\documentclass[prc,aps,twocolumn,float,floatfix,
               nofootinbib,superscriptaddress,byrevtex]{revtex4}
\usepackage{amssymb}
\usepackage{amsmath}
\usepackage{amsfonts}
\usepackage{epsfig}

\renewcommand{\vec}[1]{\mathbf{#1}}
\newcommand{\nn}{\nonumber}
\newcommand{\etal}{\emph{et al.}}
\newcommand{\nuc}[2]{$^{#1}$\textrm{#2}}

\begin{document}
\title{Configuration mixing of angular-momentum and particle-number projected\\
       triaxial HFB states using the Skyrme energy density functional}
\author{Michael Bender}
\affiliation{Service de Physique Nucl\'eaire Th\'eorique,
             Universit\'e Libre de Bruxelles, C.P. 229, B-1050 Bruxelles,
             Belgium}
\affiliation{Institute for Nuclear Theory,
             University of Washington,
             Box 351550, Seattle, WA 98195,
             USA}
\affiliation{Physics Division,
             Argonne National Laboratory,
             9700 S. Cass Avenue,
             Argonne, IL 60439,
             USA}
\affiliation{National Superconducting Cyclotron Laboratory,
             Michigan State University,
             1 Cyclotron,
             East Lansing, MI 48824,
             USA}
\affiliation{DAPNIA/SPhN,
             CEA Saclay,
             F-91191 Gif sur Yvette Cedex,
             France}
\affiliation{Universit{\'e} Bordeaux,
             Centre d'Etudes Nucl{\'e}aires de Bordeaux Gradignan, UMR5797,
             F-33175 Gradignan, France}
\affiliation{CNRS/IN2P3,
             Centre d'Etudes Nucl{\'e}aires de Bordeaux Gradignan, UMR5797,
             F-33175 Gradignan, France}
\author{Paul-Henri Heenen}
\affiliation{Service de Physique Nucl\'eaire Th\'eorique,
             Universit\'e Libre de Bruxelles, C.P. 229, B-1050 Bruxelles,
             Belgium}
\date{5 August 2008}
%
%
\begin{abstract}
We present a method based on mean-field states generated by triaxial
quadrupole constraints that are projected on particle number and angular
momentum and mixed by the generator coordinate method on the quadrupole
moment. This method is equivalent to a seven-dimensional GCM calculation,
mixing all five degrees of freedom of the quadrupole operator and the gauge
angles for protons and neutrons. A first application to $^{24}$Mg permits
a detailed analysis of effects of triaxial deformations and of $K$ mixing.
\end{abstract}
\maketitle
%
%
\section{Introduction}

Methods based on the self-consistent mean-field approach
\cite{RMP} are up to now the only microscopic tools that can be
applied to all nuclei including the heaviest ones. However,
mean-field methods have several limitations, and the method that
we present here is part of an attempt to eliminate two of the
most penalizing presented in a series of papers . The first
limitation is due to the determination of a wave function in an
intrinsic frame of reference. Although the symmetry-breaking
mean-field approach is a very efficient and transparent way to
grasp the effect of correlations associated with collective modes
in the limit of strong correlations
\cite{Rin80a,Abe90a,Naz92a,Fra01a}, the absence of good quantum
numbers and the corresponding selection rules does not allow 
direct determination of the electromagnetic transition probabilities.
Instead, approximations have to derived based on the collective
model and these cannot cover all possible cases as they are well
justified only in the limit of large deformation. The second
limitation concerns nuclei for which a mean-field description
through a single configuration breaks down because several
configurations with different shell structure are close in energy
without being separated by a large potential barrier. Methods to
overcome these two problems have been proposed in the past,
but it is only in the last ten years that sufficient computational
resources have become available to construct and apply methods
based on realistic effective energy density functionals aiming at
eliminating both of these limitations.

Angular momentum projection \cite{Won75a,Rin80a,BlaRip} is the key
tool to transform the mean-field wave function from the intrinsic
to the laboratory frame of reference. There is no ambiguity in
determining directly electromagnetic transition probabilities when
working in the laboratory system. However, without the simplifying
assumption of axial states, the restoration of rotational symmetry
requires that rotations about three Euler angles be considered.
So far, for mean-field states with triaxial quadrupolar
deformations, this has been mostly done in the context of
phenomenological models by using small shell model spaces and often
schematic interactions
\cite{Gir69a,Gru81a,Har82a,Hay84a,Wus85a,Ena99a}. In the context
of energy density functional methods, projection on angular
momentum of mean-field states with triaxial quadrupolar
deformations has been performed at the Hartree-Fock level by using a
simplified Skyrme interaction in Ref.~\cite{Bay84a} and very
recently with a full Skyrme energy functional in
Ref.~\cite{Zdu07a}. Cranked wave functions have been projected in
both cases to approximate a variation after projection on the
angular momentum \cite{Kam68a,Bec70a,Man75a}, but pairing
correlations were not included.

To solve the problem of energy surfaces that are soft with
respect to a collective degree of freedom, one has to introduce
fluctuations in this collective degree of freedom into the 
ground-state wave function using the generator coordinate method (GCM)
\cite{Oni66a,Won75a,BlaRip,Rin80a}. Again, many studies used small
shell model spaces and schematic interactions, but there is also a
quite a large body of work starting from mean-field methods based
on effective interactions in the full model space of occupied
single-particle states and with inclusion of pairing correlations.
There have been several applications~\cite{Bon90a,Bon91a,Hee93a}
that deal with the intrinsic quadrupole mode including triaxial
deformations, which is in most cases the dominant low-lying
collective excitation mode in nuclei. There also have been studies
of octupole modes \cite{Bon91b,Ska93a,Zbe06a} and their coupling
to the quadrupole mode \cite{Ska93b,Mey95a}, as well as
investigations of fluctuations in pairing degrees of freedom
\cite{Mey91a,Hee01a,Ben07a}. All the studies mentioned so far have
in common the fact that simultaneous symmetry restoration, if
performed at all, is limited to numerically inexpensive modes,
such as particle number or parity, whereas angular momentum is not
restored.

A simultaneous treatment of angular momentum projection and of
fluctuations with respect to triaxial quadrupole deformation
requires that five degrees of freedom be considered. The most
transparent representation consists of three Euler angles defining
the relative orientation of the intrinsic major axis frame in the
laboratory frame and two independent degrees of freedom
characterizing the intrinsic deformation (e.g., through the
coordinates $\beta$ and $\gamma$ introduced by {\AA}.~Bohr). In the
context of the GCM, however, this has so far only been done in
highly schematic models, by using either a single-$j$ shell
\cite{Bur95a} or a very small shell model space and a schematic
interaction \cite{Ena01a,Ena02a}.

A first step toward the simultaneous treatment of these five
degrees of freedom has been carried out recently with the GCM
mixing of angular-momentum projected quadrupole-deformed axial
mean-field states. This scheme is nowadays routinely applied by
several groups, using  Skyrme energy density functionals
\cite{Val00a,Ben03b,Ben04a}, the density-dependent Gogny force
\cite{Rod02a,Rod02b,Egi04a}, or a relativistic point-coupling
model \cite{Nik06a,Nik06b}. In all of those cases, intrinsic
triaxiality is neglected and two of the three Euler angles can be
treated analytically.

All the studies mentioned so far that consider angular momentum
projection have in common the fact that this operation is
performed after variation. This has several drawbacks
\cite{Rin80a}, in particular when working with time-reversal
invariant mean-field states. The alternative, variation after
projection, is computationally very demanding. The only standard
method where variation after angular-momentum projection is
considered (together with restoation of $N$, $Z$, and parity) is
the VAMPIR/MONSTER approach \cite{Schm87a,Schm04a} which is
confined to a very small shell model space and a shell model
effective Hamiltonian. In this framework, the intrinsic dynamics
is not described in terms of deformation degrees of freedom, but
rather in terms of quasiparticle excitations.

An approximate scheme to describe five-dimensional quadrupole
dynamics in a fully microscopic framework has been set up long
ago. The idea is to construct the input of the Bohr-Hamiltonian
microscopically \cite{Rin80a,Bon90a,Rei87a}. This is an approximation 
to the full five-dimensional GCM as it replaces the nonlocal kernels
entering the Hill-Wheeler-Griffin equation for nonorthogonal
weight functions by local potentials and mass parameters in a
collective Schr{\"o}dinger equation for orthogonal collective wave
functions, but it also
allows one to incorporate improved moments of inertia at a
moderate cost. This approach has routinely been used for a long
time in the framework of the density-dependent Gogny force
\cite{Gir78b,Lib99a}, and it has also been recently set up for Skyrme
interactions \cite{Pro04a}.

In this paper we present a first step toward a microscopic
treatment of triaxial quadrupole dynamics and of angular momentum
projection in the context of nuclear energy density functional
methods using the full space of occupied single-particle states.
Th emethod employed generalizes that described in
Refs.~\cite{Val00a,Ben03b,Ben04a} by allowing for the breaking of
axial symmetry of the mean-field wave functions.
%
%
\section{The Method}
\label{sect:method}

For a given nucleus, the calculations are performed in three
steps. First, a nonorthogonal basis of self-consistent mean-field
states is generated with constrained Hartree-Fock-Bogoliubov
(HFB) calculations. Second,
angular-momentum and particle-number projected matrix elements
between all mean-field states are calculated. Third, these matrix
elements are used in a configuration-mixing calculation to
determine the correlated ground state, the spectrum of excited
states, and the transition moments

The only inputs to the calculation are the proton and neutron
number and the parameters of a universal energy functional. The
latter are taken from the literature and obtained from a global
fit to nuclear properties aiming at the description of the entire
chart of nuclei, without local fine tuning. One consistency
requirement of our method is that the same effective interaction
is used to generate the mean-field states and to calculate the
configuration mixing. We chose a Skyrme energy functional
supplemented by the Coulomb interaction in the mean-field channel,
together with a zero-range, density-dependent functional in the
pairing channel.

We will give here, and in the Appendix, only a sketch of the
ingredients of the method. We postpone a detailed description to a
future paper on the generalization of the code to time-reversal
invariance breaking triaxial states.
%
%
\subsection{The mean-field basis}
\label{sect:mf}

The HFB equations are solved self-consistently with the two-basis
method of Ref.~\cite{Gal94a}. The quadrupole moment is constrained
through two coordinates $q_1$ and $q_2$, which are related to the
usual mass quadrupole deformations $\beta$ and $\gamma$ of the
Bohr Hamiltonian through the relations \cite{Bon05a}:
\begin{eqnarray}
\label{eq:constraint}
q_1
& = & Q_0 \, \cos (\gamma) - \frac{1}{\sqrt{3}} \, Q_0 \, \sin (\gamma)
      \\
q_2
& = & \frac{2}{\sqrt{3}} \, Q_0 \, \sin (\gamma)
\, .
\end{eqnarray}
with
\begin{equation}
\label{eq:beta}
\beta_2
= \sqrt{\frac{5}{16\pi}} \; \frac{4\pi Q_0}{3 R^2 A}
\end{equation}
where $R = 1.2 \, A^{1/3}$ fm. A mesh of positive values for $q_1$
and $q_2$ covers the entire first sextant of the $\beta$-$\gamma$
plane. The mean-field equations are solved on a three-dimensional
mesh~\cite{{Bon05a}}, with the total nuclear density imposed to be
symmetric with respect to three planes. Thus, all odd-$l$ moments of
the density identically vanish whereas the even moments, which are
not constrained, are fully taken into account and take the values
that minimize the energy. All the wave functions that we consider
are, therefore, restricted to have positive parity.

For the results described in the following, the mean-field wave functions 
are required to be time-reversal invariant, such that the collective
coordinates can be limited to one sextant of the $\beta$-$\gamma$
plane. The five other sextants correspond to one or several
permutations of the principle axes of inertia. After projection
and configuration mixing, all of them are equivalent, although
intermediate (nonobservable) quantities might differ. This
equivalence will be used as a test of the numerical implementation
of our method of Sec.~III.
%
%
\subsection{Projection}

Restoring symmetries allows to extract states with good quantum
numbers from the mean-field states and provides a clean framework
to use selection rules for electromagnetic transitions.
Eigenstates of the particle-number operator are obtained in the
same way as previously~\cite{Ben03b,Ben04a}. From a technical
point of view, it is crucial to restore this symmetry when mixing
wave functions that are not eigenstates of the particle-number
operators. Since those wave functions have only the right mean
particle number, their transition matrix elements will contain
nondiagonal contributions in the number of particles. One can
estimate easily the error that this symmetry breaking brings to
energies. The binding energy per nucleon of most nuclei is between
7 and 8~MeV/u and a deviation of the particle number of a mixed
state as small as 0.1 nucleons would affect its energy by several
hundreds of keV. Particle-number projection removes this ambiguity
\cite{Hee93a}. A technically simpler approximate method would be
to introduce a constraint on the average value of the particle
number $N_0$ of the mixed states in the Hill-Wheeler-Griffin
equation \cite{Bon90a,Rod02b}, but since the same value for this
constraint has to be used for all the eigenstates to maintain
their orthogonality, the error on the mean number of particles
will vary from one state to another. We always project on the
proton and neutron numbers imposed on the mean-field wave
functions by Lagrange multipliers in the mean-field equations. We,
therefore, drop any reference to particle numbers for the sake of
simple notation.

Angular momentum projection is significantly more complicated when
triaxial mean-field states instead of axial ones are considered.
Eigenstates of the total angular momentum operator in the
laboratory frame, $\hat{J}^2$, and its projection on the $z$ axis,
$\hat{J}_z$, with eigenvalues \mbox{$\hbar^2 J (J+1)$} and $\hbar M$,
respectively, are obtained by applying the operator
\begin{equation}
\label{eq:PJ}
\hat{P}^J_{MK}
= \frac{2J+1}{16 \pi^2}
  \! \int_{0}^{4\pi} \! \! d\alpha
  \! \int_0^\pi \! \! d\beta \; \sin(\beta)
  \!  \int_0^{2 \pi} \! \! d\gamma \;
  \mathcal{D}^{J \, \ast}_{MK} \, 
  \hat{R} 
\, ,
\end{equation}
where \mbox{$\hat{R} = e^{-i\alpha \hat{J}_z} \, e^{-i\beta \hat{J}_y} \,
e^{-i\gamma \hat{J}_z}$} is the rotation operator and
$\mathcal{D}^{J}_{MK} (\alpha,\beta,\gamma)$ a Wigner function
\footnote{
Alternatively, the integration intervals can be chosen
as $\alpha \in [0,2\pi]$, $\beta \in [0,\pi]$, and $\gamma \in
[0,4\pi]$. In any case, for systems with integer $J$ values as
discussed here, the integration over $[2\pi,4\pi]$ of either
$\alpha$ or $\gamma$ gives just a factor of 1.
}
\cite{Var88a}. $\hat{P}^J_{MK}$ picks the component with angular momentum
projection $K$ along the intrinsic $z$ axis. The projected state
is then obtained by summing over all $K$ components of the
mean-field state $| q \rangle$,
\begin{eqnarray}
\label{eq:JMkappaq}
| J M \kappa q \rangle
& = & \sum_{K=-J}^{+J} f^{J}_{\kappa} (K) \; \hat{P}^J_{MK} \, | q \rangle
      \nn \\
& = & \sum_{K=-J}^{+J} f^{J}_{\kappa} (K) \; | J M K q \rangle
\end{eqnarray}
with weights $f^{J}_\kappa (K)$ determined by minimizing the
energy~\cite{Won75a,Rin80a} (see the following). The number of
values that the index $\kappa$ can take is restricted by the
symmetries of the mean-field states~\cite{Ena99a} (signature with respect
to $x$ and time-reversal invariance) to $J+1$ for even $J$ values
and $J-1$ for odd $J$ values.

Note that $\hat{P}^{J}_{KM}$ is not a projector in the strict
mathematical sense, but it has the properties
$\hat{P}^{J}_{KM} \hat{P}^{J'}_{M' K'} = \delta_{J J'}
\, \delta_{M M'} \, \hat{P}^{J}_{K K'}$ and
$(\hat{P}^{J}_{KM})^\dagger = \hat{P}^{J}_{MK}$ \cite{Cor71a}.
%
%
\subsection{Mixing of states with different deformations}

The fluctuations of the intrinsic deformation and the resulting
spreading of the nuclear states in the $\beta$-$\gamma$ plane can
be described by a superposition of angular-momentum projected
states using the two intrinsic quadrupole degrees of freedom $q$
as generator coordinates. Taking into account that one has also to
mix all $K$ values for each deformation, one obtains a resulting wave
function given by
\begin{equation}
\label{eq:J:GCM}
| J M \nu \rangle = \sum_{q} \sum_{K} F^{J}_{\nu} (K,q)
  \, | J M K q \rangle
\, ,
\end{equation}
This corresponds to the discretized version of the GCM.
$F^{J}_{\nu} (K, q)$ is a weight function of the $K$ components of
the angular-momentum projected states of intrinsic deformation
$q$, determined from
\begin{equation}
\frac{\delta}{\delta F^{J \, \ast}_{\nu} (K,q)}
\frac{\langle J M \nu | \hat{H} | J M \nu \rangle}
     {\langle J M \nu | J M \nu \rangle}
= 0
\, ,
\end{equation}
which leads to the Hill-Wheeler-Griffin (HWG)
equation \cite{Hil53a,Gri57a}
\begin{eqnarray}
\label{eq:HWG:full}
\lefteqn{
\sum_{q'} \sum_{K' }
\big[  \mathcal{H}_J(K, q; K', q')
}  \nonumber \\
&  & \quad
      - E^{J}_{\nu} \, \mathcal{I}_J (K, q; K', q') \big]
     \; F^{J}_{\nu} (K',q')
= 0
\end{eqnarray}
with the energy and norm kernels
\begin{eqnarray}
\label{eq:HWG:kernels}
\mathcal{H}^J(K, q ; K',q')
& = & \langle J M K q | \hat{H} | J M K' q' \rangle \\
\mathcal{I}^J(K, q ; K', q') & = & \langle J M K q | J M K' q'
\rangle \, .
\end{eqnarray}
For the sake of simple notation, we have introduced the method
using a Hamilton operator $\hat{H}$. We shall comment
on the procedure used to calculate the Hamiltonian kernels from an
energy density functional, and discuss some problems this may cause
in Sec.~\ref{sect:MREDF}.

The dimension of the variational space is considerably increased
by introducing triaxial deformation. One cannot limit the
angular momentum projection to $K = 0$, and several $K$ components
for each deformation have to be mixed. This has to be done with
some care because of the high redundancy of the GCM bases, in the
$K$ space for a given deformation on the one hand and also for the
whole set of deformations on the other hand. For an efficient
elimination of redundant states we have chosen a four-step
procedure. We start from a basis $| J M K q \rangle$. First, we
restrict the subspace for each deformation $q$ and angular
momentum $J$ by diagonalizing the norm kernel $\mathcal{I}^J(K, q
; K',q)$, rewritten in its significant subspace~\cite{Ena99a}, and
neglecting eigenstates with negligible eigenvalues, typically
smaller than $10^{-3}$. Second, we solve the HWG equation in $K$
space within this new basis. The solutions $| J M \kappa q
\rangle$ are labeled by an index $\kappa$,
Eq.~(\ref{eq:JMkappaq}). The number of values that $\kappa$ can
take for a given $J$ depends on the deformation. It is just one
for axial states and even $J$ values and is usually larger for
triaxial deformation. In a third step, we transform all matrix
elements to the new basis $| J M \kappa q \rangle$ and diagonalize
the norm matrix $\mathcal{I}^J(\kappa, q ; \kappa',q')$ in the
combined $\kappa$ and $q$ space. Once more, only significant
eigenvectors of the norm matrix are retained. Finally, the
Hamiltonian kernel is diagonalized in the basis of norm
eigenstates to construct the weight functions $F^{J}_{\nu}
(\kappa,q)$. The final eigenstates $| J M \nu \rangle$,
Eq.~(\ref{eq:J:GCM}), mix all $K$ and $q$ values. They are then
used to calculate all observables and transition moments.

Note that our method is not restricted to the choice of the
triaxial quadrupole moment as a generator coordinate. This choice
is suggested by the importance of quadrupole correlations in
nuclei, but any other collective variable associated with an even
multipole moment could be additionally considered, at the expense
of having a larger mean-field basis. Inncluding odd multipoles, in
particular octupole deformations, would require a relatively
simple generalization of the codes but cannot be considered at
present. However, some modes not related to a shape degree
of freedom, such as pairing vibrations, can be included without
modification of the numerics, as done earlier without angular
momentum projection in Refs.~\cite{Mey91a,Hee01a,Ben07a}.
%
%
\subsection{Electromagnetic matrix elements}
\label{sect:trans}

Once the HWG equation is solved and the weight functions
$F^{J}_{\nu} (\kappa,q)$ are known, the expectation values and
transition moments of other observables $\hat{O}$ can be determined
starting from the kernels
$\langle J M K q | \hat{O} | J M K' q' \rangle$
of the corresponding operators. Some of them provide a good
test of the accuracy of symmetry restoration, which can be used to
determine a sufficient number of points for projection. The matrix
elements of the proton and
neutron number operators, for example, are equal to the required
values with an absolute deviation lower than $10^{-8}$, and they have a
dispersion lower than $10^{-8}$ when using a nine-point formula for
the particle-number projection.

The symmetries of the unrotated time-reversal invariant HFB states
(time reversal and triaxiality) permit the reduction of the
integration intervals for Euler angles to $\alpha \in [0,\pi/2]$,
$\beta \in [0,\pi/2]$, and $\gamma \in [0,\pi]$ (i.e., 1/16 of the
full $8\pi^2$ integration volume for systems with integer spin).
The integrals over $\alpha$ and $\gamma$ are discretized with a
trapezoidal rule, while for $\beta$ we employ Gauss-Legendre
quadrature. The number of points in these intervals used for the
results reported in the following is 6 for $\alpha$, 18 for $\beta$, 
and 12 for $\gamma$, which corresponds to 24, 36, 24 points in the 
full $8\pi^2$ integration volume necessary for integer $J$ values. With
this discretization, the calculated expectation values of the
angular momentum operators $\hat{J}^2$ and $\hat{J}_z$ are
accurate with an error of the order $10^{-4}$ for the values of
$J$ discussed here.

Since the restoration of rotational symmetry provides wave
functions in the laboratory frame of reference, one can compute
directly the expectation values and transition matrix elements of
electromagnetic operators. Besides mean-square radii and $E0$ transition
moments, we calculate spectroscopic quadrupole moments and $B(E2)$
values as well as magnetic moments and $B(M1)$ values.
By combining projection and variational GCM mixing of states
with different mass quadrupole moments, the $B(E2)$ transition
moments for in-band and out-of-band transitions take the form
\begin{eqnarray}
\lefteqn{
B (E2; J_{\nu'}' \to J_\nu)
} \nn \\
& = & \frac{e^2}{2J'+1}
      \sum_{M =-J }^{+J }
      \sum_{M'=-J'}^{+J'}
      \sum_{\mu=-2}^{+2}
      | \langle J M \nu \big| \hat{Q}_{2 \mu} | J' M' \nu' \rangle \big|^2
      \nonumber \\
&   &
\end{eqnarray}
and the spectroscopic quadrupole moments are given by
\begin{equation}
Q_s (J_\nu)
= \sqrt{\frac{16\pi}{5}} \,
  \langle J M=J \; \nu | \hat{Q}_{2 0} | J M=J \; \nu \rangle
\, .
\end{equation}
The matrix elements of the electric quadrupole moment operator
$Q_{2\mu} = e \,\sum_p r_p^2 \, Y_{2\mu}(\Omega_p)$ that enter the
$B(E2)$ values and $Q_s$ are calculated by using point protons with
their bare electric charge $e$. This approach is justified by the
following considerations. First, there is no empirical evidence
that electric moments are significantly modified by the in-medium
effects that are absorbed into the effective energy functional
but ignored for all other observables. Second, we use the entire
space of occupied single-particle states without assuming an inert
core. Third, the projected GCM, as a "horizontal expansion"
\cite{Don89a} of large-amplitude dynamics, can be formulated such
that the calculation of matrix elements does not contain a sum
over unoccupied states \cite{Bon90a,Hee93a,FloThesis}. The absence
of effective charges is an important feature of a method aiming at
a universal description of nuclei. The (at least approximate)
folding with the intrinsic charge distribution of protons and
neutrons to construct a charge density is important for radii
\cite{RMP}, but it plays no role for multipole moments when accepting
the current precision of our approach.

Both the $B(E2)$ and $Q_s$ values scale with mass and angular
momentum. With triaxial shapes, however, on can no longer
define a dimensionless intrinsic deformation
$\beta_2$ that corresponds to a single $Q_s$ or a single $B(E2)$
value, as can be done in the axial case \cite{Ben03b}, as one
needs two independent observables (i.e., that are not within the
same band) to determine the deformation $\beta_2$ and the angle
$\gamma$ \cite{Cli86a}.

Similarly, magnetic moments are given by
\begin{equation}
\mu (J_\nu)
= \langle J, M=J, \nu | \hat{\mu}_{1 0} | J, M=J, \nu \rangle
\end{equation}
and $B(M1)$ values are obtained as
\begin{eqnarray}
\lefteqn{
B (M1; J_{\nu'}' \to J_\nu)
} \nn \\
& = &\frac{3}{4 \pi} \frac{1}{2J'+1} \!
      \sum_{M =-J }^{+J }
      \sum_{M'=-J'}^{+J'}
      \sum_{\mu=-1}^{+1}
      \big| \langle J M \nu | \hat{\mu}_{\mu} | J' M' \nu' \rangle \big|^2
      \nonumber \\
&   &
\end{eqnarray}
The magnetic dipole operator $\hat{{\boldsymbol{\mathbf{\mu}}}}$
entering both expressions is given by
\begin{equation}
\hat{{\boldsymbol{\mathbf{\mu}}}}
=   g_{\ell,p} \hat{\vec{L}}_p
  + g_{s,p} \hat{\vec{S}}_p
  + g_{s,n} \hat{\vec{S}}_n
\, ,
\end{equation}
where $\hat{\vec{L}}_t$ and $\hat{\vec{S}}_t$ are the total
orbital and spin operators for protons and neutrons, $t=p,n$, and
$g_{\ell,p} =  1 \, \mu_N$, $g_{s,p} = 5.585 \, \mu_N$, and
$g_{s,n} = - 3.826 \, \mu_N$ are the $g$ factors of protons and
neutrons in units of the nuclear magneton $\mu_N$. Unlike the
electric moments, the spin contribution to the magnetic moments is
modified by the physics of short-range correlations which is
resummed into the energy functional but not explicitly considered
in the wave functions. The spin $g$ factors are often quenched by
an empirical factor to mock up these effects. In the present
paper, we start with time-reversal invariant mean-field states for
which the spins of conjugated states exactly cancel each other.
Although this is no longer the case after mixing such states,
particularly when projecting on angular momentum, the
angular-momentum projected total spin $\langle J \, \mbox{$M$=$J$}
\, \nu | \hat{S}_z | J \, \mbox{$M$=$J$} \, \nu \rangle$, remains
a small correction of a few percent to the dominating collective
contribution from orbital angular momentum when calculating the
$z$ component of total angular momentum in the laboratory frame,
$\langle J \, \mbox{$M$=$J$} \, \nu | \hat{J}_z | J \,
\mbox{$M$=$J$} \, \nu \rangle = \hbar J$. This has two
consequences. First, we can ignore the subtlety of in-medium
corrections to the spin $g$ factors for time being. However, when
projecting and mixing cranked states at high spin or
$n$-quasiparticle states that might have significant open spin,
this point will deserve more attention. Second, the magnetic
moments calculated for the projected and mixed states discussed in
the present paper will be very close to the expectation value of
the proton orbital angular momentum $\langle J \, \mbox{$M$=$J$}
\, \nu | \hat{L}_{z,p} | J \, \mbox{$M$=$J$} \, \nu \rangle$ in
nuclear magnetons.

%
%
\subsection{Evaluation of kernels}
\label{sect:kernels}

The matrix elements that are needed to solve the HWG equation and
to calculate the electromagnetic properties of the GCM eigenstates
are matrix elements of tensor operators $\hat{T}_{\lambda \mu}$ of
rank $\lambda$ between particle-number and angular-momentum
projected states $ \langle q | \hat{P}^N \hat{P}^Z \hat{P}_{KM}^J
            \hat{T}_{\lambda \mu}
            \hat{P}_{M'K'}^{J'} \hat{P}^Z \hat{P}^N | q' \rangle
$. As at present we consider only operators that commute with the
particle-number operator, the particle-number projector for the
left state can be commuted with $\hat{T}_{\lambda \mu}$ and one
only has to project the right state on particle number. The
angular-momentum projection operator, however, does not commute with
tensor operators of rank different from zero. Still, the commutator can
be evaluated using angular-momentum algebra to obtain matrix
elements with an angular-momentum projection operator acting on
the left state only:
\begin{eqnarray}
\lefteqn{
\langle q | \hat{P}_{KM}^J \hat{T}_{\lambda \mu} \hat{P}_{M'K'}^{J'}
| q' \rangle
} \nn \\
& = & \frac{2J'+1}{2J+1} \; (J' \lambda J | M' \mu M )
      \sum_{k=-J}^{+J} (J' \lambda J | K', K' - k, k) \;
      \nn \\
&   & \quad \times
     \langle q | \hat{P}^{J}_{K k} \; \hat{T}_{\lambda, K' - k} | q' \rangle
     \, .
\end{eqnarray}
 For a scalar operator $\hat{T}_{00}$, this simplifies to
\begin{equation}
\langle q | \hat{P}_{KM}^J \, \hat{T}_{00} \,
            \hat{P}_{M'K'}^{J'} | q' \rangle
= \delta_{J J'} \delta_{M M'}
  \langle q | \hat{P}^{J}_{K K'} \; \hat{T}_{00} | q' \rangle
\, .
\end{equation}
As a consequence, we have to evaluate matrix elements of the form
\begin{equation}
\label{eq:meLR}
\langle q | \hat{R}^\dagger (\alpha,\beta,\gamma) \, \hat{T}_{\lambda \mu} \,
           e^{i \phi_n \hat{N}} \, e^{i \phi_z \hat{Z}} | q' \rangle
= \langle \text{L} | \hat{T}_{\lambda \mu} | \text{R} \rangle
\end{equation}
where $| q \rangle$ and $| q' \rangle$ are HFB states with
possibly different intrinsic deformations, and $| \text{L}
\rangle$ and $| \text{R} \rangle$ are the HFB states rotated in
coordinate and gauge space, respectively. Key elements of
the evaluation of these kernels are outlined in the
Appendix~\ref{app:representation}.

%
%
\subsection{Configuration mixing using energy functionals}
\label{sect:MREDF}

\begin{figure}[t!]
\includegraphics{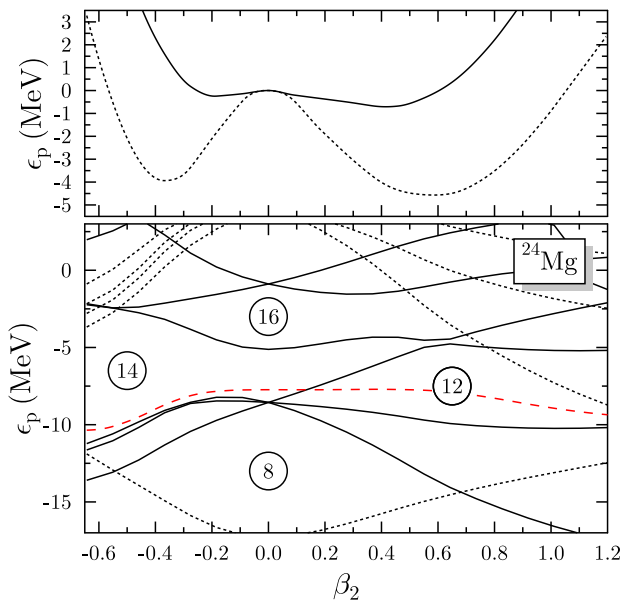}
\caption{\label{fig:nilsson:ax}
(Color online)
Lower panel: Nilsson diagram of the eigenvalues of the proton single-particle
Hamiltonian for axial shapes. Solid lines denote levels with positive
parity, dotted lines levels with negative parity, and the red
dashed line the Fermi energy.
Upper panel: Particle-number projected (solid line) and particle-number
and angular-momentum $J=0$ projected (dotted line) axial deformation
energy curves.
}
\end{figure}

For the sake of simplicity, we have outlined the projected
GCM using a many-body Hamilton operator. In practice, however, an
energy density functional is used in a form postulated by analogy
with the Hamiltonian case~\cite{Bon90a,Rei94a}. In this procedure,
the multi-reference (MR) energy density functional (EDF) is
obtained by replacing the density matrices in the single-reference
energy functional by the transition density matrices as they
appear when transition matrix elements of a Hamilton operator are
evaluated with the generalized Wick theorem
\cite{Bal69a,Rin80a,Bon90a}. The terms in the Skyrme EDF that
contain time-odd densities are treated as described in Appendix~C
of Ref.~\cite{Bon90a} and give a small contribution to the
nondiagonal kernels. The density-dependent terms are generalized
by using the standard prescription that the density is replaced by
the transition density in the density-dependent terms. This is the
only prescription that guarantees various consistency requirements
of the energy functional \cite{Rod02b,Rob07a}. However, this
procedure may lead in some cases to problems that have become
evident recently~\cite{Ang01a,Dob07a}. Discontinuities or even 
divergences of the energy kernel may indeed appear as a function of 
deformation. A regularization scheme has been proposed
\cite{Lac08a} and applied to a simple case of MR-EDF calculations
\cite{Ben08a}. However, it cannot be applied to the standard form
of the Skyrme and Gogny interactions \cite{Dug08a}. We leave this
problem to be addressed in the future and stick to a standard
Skyrme energy functional, as done in our earlier studies.

%
%
\section{$^{24}$Mg as an illustration}

The main features of our method are exemplified by a calculation
for \nuc{24}{Mg}. We use the parametrization SLy4 \cite{Cha98a} of
the Skyrme+Coulomb energy density functional for the particle-hole
channel of the effective interaction, supplemented by a zero-range
density-dependent pairing energy functional~\cite{Rig99a}. A soft
cutoff around the Fermi energy is used when solving the HFB
equations as introduced in Ref.~\cite{Bon85a}.

Figure~\ref{fig:nilsson:ax} provides as a reference some key
results from an \emph{axial} calculation as a function of the
dimensionless mass quadrupole deformation given through
Eq.~(\ref{eq:beta}). In this figure, oblate shapes are labeled by
negative values of $\beta_2$. On the top part we compare the
variation of the energy with deformation obtained by projecting on
particle numbers $N$ and $Z$ only (solid line) and by projecting
on particle number and on $J=0$ (dotted line). In the bottom part,
the Nilsson diagram of eigenvalues of the proton single-particle
Hamiltonian is shown . The Nilsson diagram for neutrons is nearly
identical to the one for protons, except for a global shift owing to
the absence of the Coulomb potential. The projection on $J=0$
significantly increases the energy gain from deformation. The
mean-field configuration corresponding to the minimum  also has a
significantly larger prolate deformation after projection than
before. After projection, the axial minimum corresponds to the
intrinsic deformation where the $Z=N=12$ gap in the Nilsson
diagram is largest. The Nilsson diagram indicates also that there
are only very few level crossings in the interesting region of
deformations.

%
%
\subsection{Triaxial energy maps}

\begin{figure}[t!]
\includegraphics[width=6cm]{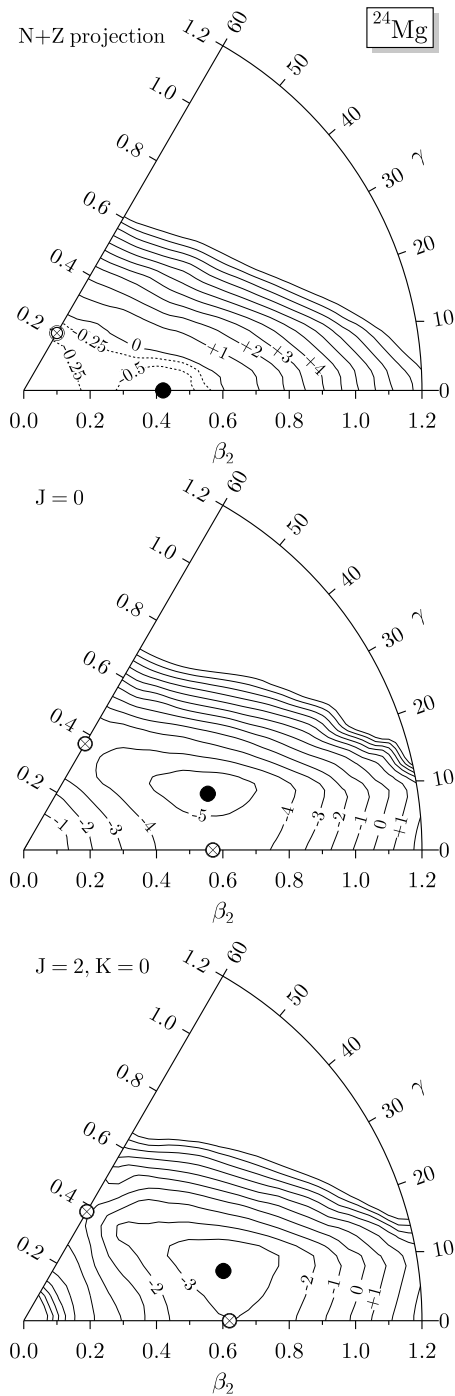}
\caption{\label{fig:map:J00}
Triaxial quadrupole energy maps obtained by projecting mean-field
configurations on $N$ and $Z$ only (top) and also either on $J=0$
(middle) or on $J=2$ and $K=0$. In the latter case, the $z$ axis
is chosen as symmetry axis. Filled circles denote minima; the
$\otimes$ symbols denote saddle points. All energies are normalized with
respect to the energy of the particle-number projected spherical
state. }
\end{figure}

Three ($\beta$, $\gamma$) deformation energy surfaces for
\nuc{24}{Mg} are displayed in Fig.~\ref{fig:map:J00}. All energies
are normalized to the energy of the spherical configuration. A
projection on particle numbers $N$ and $Z$ is performed for all
three panels; it is combined with a projection on the angular
momentum $J=0$ in the middle panel and on $J=2$ and $K=0$ on the
lower one. Several comments on the interpretation of the
deformation energy surfaces are in order. First, the coordinates
$\beta_2$ and $\gamma$ are not related to any matrix element in
the angular-momentum projected energy surfaces but merely provide
a label for the intrinsic state fromwhich the projected energy is 
obtained. All states projected on $J=0$ are spherical in the
laboratory frame. Second, neither the mean-field nor the projected
states are orthogonal, such that the energy maps do not and cannot
represent the actual metric, which is related to the inverse of
the respective overlap matrix. Finally, the $J=2$, $K=0$ surface
depends on the orientation chosen for the principal axes of the
nucleus. The quadrupole moment along the $z$ axis is the largest
one in Fig.~\ref{fig:map:J00}. We shall show later that the $K=0$
results have no obvious interpretation for other choices.

Starting with the energy surface without angular-momentum projection, 
one sees that the mean-field ground state corresponds to an axial
prolate deformation, more bound than the spherical configuration
by about 700~keV. Comparing with the axial energy curve given as a
solid line in the upper panel of Fig.~\ref{fig:nilsson:ax}, one
notices that the prolate minimum is indeed a true one but
that the oblate extremum seen in Fig.~\ref{fig:nilsson:ax} is a
saddle point in the $\gamma$ direction. A qualitatively similar
deformation energy surface was found in HF+BCS calculations with
the Skyrme interaction SIII in Ref.~\cite{Bon87a} and in HFB
calculations with the Gogny force D1 in Ref.~\cite{Gir83a}.

The projection on $J=0$ favors triaxial configurations: The lowest
energy is obtained for a triaxial mean-field configuration with
$\gamma\approx 16^\circ$ and a value for $\beta_2$ around 0.6,
similar to that of the axial mean-field state giving the lowest
$J=0$ projected energy. Before projection, these two states were
separated by 2.2~MeV. The energy gain for the triaxial point is
3.1~MeV larger than that of the axial point. This difference is
thus large enough to compensate for the difference in energy
between the mean-field configurations. However, this result has to
be taken \emph{cum grano salis} as the deformation has a limited
meaning after projection. In particular, the states resulting from
the projection on the same $J$ value of mean-field configurations
corresponding to different quadrupole moments may have a large
overlap. The  states with the lowest energy obtained after
projection of axial and triaxial configurations have an overlap
close to 0.9.

\nuc{24}{Mg} is one of the few light nuclei with a deformed
mean-field ground state when pairing correlations are taken into
account. In this small system, however, the static quadrupole
correlation energy (i.e., the deformation energy of the mean-field
ground state) is much smaller than the dynamical quadrupole
correlation energy obtained from projection and mixing of states
with different intrinsic shapes; this is a general feature of
light nuclei with $A < 100$ \cite{Ben05a}.

\begin{figure*}[t!]
\includegraphics{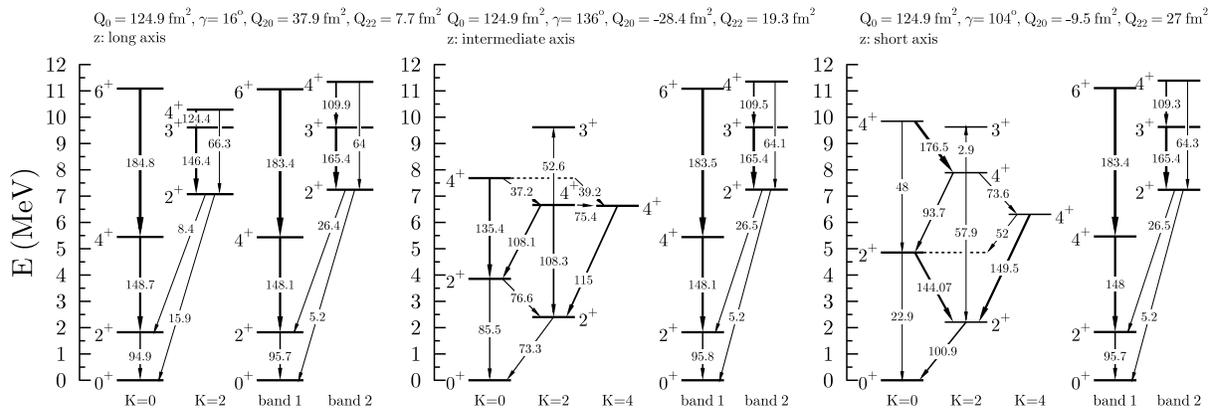}
\caption{\label{fig:orientation}
Excitation spectra and selected $B(E2)$ values after decomposition
of the mean-field state that gives the minimum of the $J=0$ projected
energy surface into angular-momentum projected components $| J M K q \rangle$
(left) and after $K$ mixing into $| J M \kappa q \rangle$ (right) for
three different orientations of the same triaxial state in the intrinsic
frame.
}
\end{figure*}

The projection on $J=2$, $K=0$ leads to an energy map whose
interpretation is more difficult than that for $J=0$. The
orientation of the nucleus that is chosen gives to this energy
map some meaning in the interpretation of the first $2^+$ level,
whereas the second $2^+$ state belongs mainly to the map
corresponding to $K=2$. The topography of the $J=2$, $K=0$ map is
quite similar to, but slightly flatter than, that of the $J=0$
energy map.

We will not show here the deformation energy surfaces for higher
$J$ and $K$ values. When starting with time-reversal invariant
mean-field states, the angular-momentum projected energy surfaces
for all even values of $J$ and $K \neq 0$ are infinite at $\gamma
= 0^\circ$, and degenerate with the $K=0$ surface of the same $J$ for
$\gamma = 60^\circ$, which we will exemplify for one oblate
configuration in the next section. As a consequence, $K$ mixing
becomes usually much stronger when going from prolate to oblate
shapes with increasing $\gamma$. For odd values of $J$, there is
no $K=0$ component, and the energy surfaces for finite $K$ are
infinite at both $\gamma = 0^\circ$ and $\gamma = 60^\circ$.

These findings are consistent with what has been seen before. Many
even-even nuclei have a quadrupole-deformed mean-field ground
state with a substantial energy gain that might be as large as
25~MeV in very heavy nuclei \cite{Ben05a}. The corresponding shape
in the intrinsic frame, however, remains axial in most cases.
Triaxial ground-state deformation is rare at the mean-field level,
and the additional energy gain from nonaxial deformation is
usually quite small compared to the deformation energy of the
axial minimum or saddle in the same nucleus
\cite{Gir78a,Gir83a,Red88a,Mol06a,Guo07a}. Triaxiality, however,
plays an important role for high-spin states when described in
mean-field approaches \cite{Szy83a,Sat05a}. The situation changes
when going "beyond the mean field". It is well known that
projection on good quantum numbers favors the breaking of the
corresponding symmetries in the underlying mean-field states: For
a nucleus with a spherical minimum in the mean-field deformation
energy surface, angular-momentum projection on $J=0$ results in a
deeper minimum at slightly deformed shapes \cite{Ben05a}, as first
noticed by Dalafi \cite{Dal75a}; for a nucleus with an axial
minimum of the mean-field energy surface, the minimum of the $J=0$
angular-momentum projected energy surface is slightly shifted into
the triaxial plane, as has been demonstrated by Hayashi \etal\
\cite{Hay84a}.
%
%
\subsection{Angular-momentum projection of a single triaxial HFB state}

\subsubsection{The role of the orientation in the intrinsic frame}

\begin{table}[b!]
\caption{\label{tab:orientation}
Excitation energies after decomposition of the mean-field state that
gives the minimum of the $J=0$ projected energy surface into angular-momentum
projected components $| J M K q \rangle$ and after $K$ mixing into
$| J M \kappa q \rangle$ for three different orientations of the same
triaxial state in the intrinsic frame, corresponding to Fig.~\ref{fig:orientation}.
}
\begin{tabular}{ccrrrccrrr}
\hline\noalign{\smallskip}
& \multicolumn{4}{c}{decomposition} &  & \multicolumn{4}{c}{$K$ mixing} \\
$J$ & $K$ & $104^\circ$ & $136^\circ$ & $16^\circ$ & & $\kappa$ & $104^\circ$ & $136^\circ$ & $16^\circ$ \\
\noalign{\smallskip} \cline{2-5} \cline{7-10}\noalign{\smallskip}
  0 & 0 &  0.00 &  0.00 &  0.00 & & 1 &  0.00 &  0.00 &  0.00 \\
\noalign{\smallskip} \hline\noalign{\smallskip}
  2 & 0 &  4.86 &  3.86 &  1.83 & & 1 &  1.84 &  1.83 &  1.83 \\
    & 2 &  2.21 &  2.40 &  7.07 & & 2 &  7.24 &  7.25 &  7.24 \\
\noalign{\smallskip}\hline\noalign{\smallskip}
  3 & 2 &  9.63 &  9.62 &  9.61 & & 1 &  9.63 &  9.62 &  9.61 \\
\noalign{\smallskip}\hline\noalign{\smallskip}
  4 & 0 &  9.85 &  7.68 &  5.45 & & 1 &  5.47 &  5.44 &  5.44 \\
    & 2 &  7.89 &  6.66 & 10.29 & & 2 & 11.39 & 11.35 & 11.34 \\
    & 4 &  6.31 &  6.63 & 16.56 & & 3 & --    &  --   &   --  \\
\noalign{\smallskip}\hline\noalign{\smallskip}
  5 & 2 & 18.34 & 16.14 & 14.52 & & 1 & 14.47 & 14.42 & 14.42 \\
    & 4 & 14.85 & 14.82 & 20.53 & & 2 & --    &  --   &  --   \\
\noalign{\smallskip}\hline\noalign{\smallskip}
  6 & 0 & 15.59 & 14.82 & 11.09 & & 1 & 11.11 & 11.08 & 11.06 \\
    & 2 & 14.14 & 12.84 & 14.27 & & 2 & 17.25 & 17.38 & 17.18 \\
    & 4 & 13.75 & 11.94 & 20.58 & & 3 & --    &  --   &   --  \\
    & 6 & 11.95 & 12.94 & 29.26 & & 4 & --    &  --   &   --  \\
\noalign{\smallskip}\hline
\end{tabular}
\end{table}

Scalar quantities such as the overlap and the energy kernels of the
GCM do not depend on the orientation of the projected state in the
laboratory frame (i.e., on its angular momentum projection $M$).
Moreover, final results do not depend on the orientation of the
mean-field state in its intrinsic frame when time-reversal
invariance is preserved. This equivalence, however, is only
obtained after $K$ mixing. This provides the opportunity to analyze 
the role of $K$ mixing for the projection of states with same intrinsic
deformation but a different orientation of their principal axes. 
It also constitutes an excellent test of the numerical accuracy of 
the projection scheme that we have developed.

\begin{figure}[t!]
\includegraphics{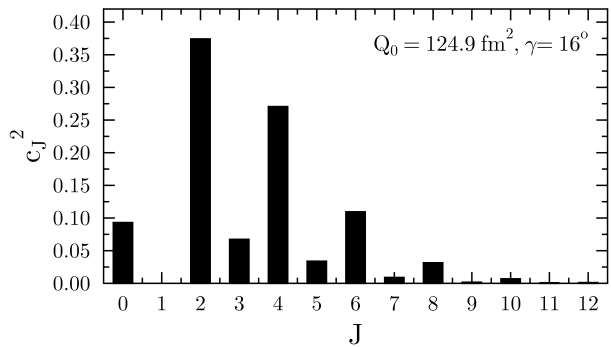}
\caption{\label{fig:decomp:j}
Decomposition of the mean-field state that gives the minimum of the
$J=0$ projected energy surface into components of same angular momentum
$J$, Eq.~(\ref{eq:cJ2}). The results obtained from the different possible
orientations in the intrinsic frame are undistinguishable.
}
\end{figure}

There are six possible ways to orient a triaxial nucleus in its
intrinsic frame in such a way that the major axes of the intrinsic
quadrupole moment coincide with the axes of the intrinsic
coordinate system. These six possibilities correspond to the six
sextants of the $\beta$-$\gamma$ plane. Owing to the particular
role played by the intrinsic $z$ axis in the angular-momentum
projector (\ref{eq:PJ}), results are invariant under exchange of
the $x$ and $y$ axes such that they give pairwise the same
decomposition $| J M K q \rangle$. It is, therefore, sufficient to
consider the intrinsic $z$ axis coinciding with the longest,
intermediate, or shortest axis of the triaxial state, without
specifying the orientation of the other two axes. The results for
excitation energies and $B(E2)$ values obtained for the three
possible orientations of the nucleus are presented in
Fig.~\ref{fig:orientation}; key numbers for excitation energies
are repeated in Table~\ref{tab:orientation}. When the intrinsic
long axis is chosen along the $z$ axis (left panel), $K$ mixing
has a small effect and a clear connection can be made between the
$K=0$ and the $K=2$ bands and the ground-state and $\gamma$ bands
after mixing. The differences between the results before and after
$K$ mixing are due to the nonorthogonality of the states $| J M K
q \rangle$ with the same $J$ and $M$ values but different $K$,
which is removed in the orthogonal basis $| J M \kappa q \rangle$.
This orthogonalization pushes up the second $4^+$ state by 1~MeV.
The dominating in-band $B(E2)$ values are similar before and after
$K$ mixing, whereas the much smaller out-of-band $B(E2)$ values
change substantially.

The situation is quite different when the triaxial mean-field
state is orientated in such a way that the $z$ axis is not the
longest one. The energies and transition probabilities obtained
before $K$ mixing have no obvious correspondence with the
$K$-mixed results. In particular, the $B(E2)$ values without $K$
mixing are of similar size for in-band and out-of-band
transitions. However, both Fig.~\ref{fig:orientation} and
Table~\ref{tab:orientation} clearly indicate the independence of
the results from the orientation of the mean-field state after $K$
mixing. Note that the $3^+$ level is not affected by $K$ mixing: as
with good time-reversal invariance,there is only one independent
nonzero $K$ component with $K = 2$.

\begin{figure}[t!]
\includegraphics{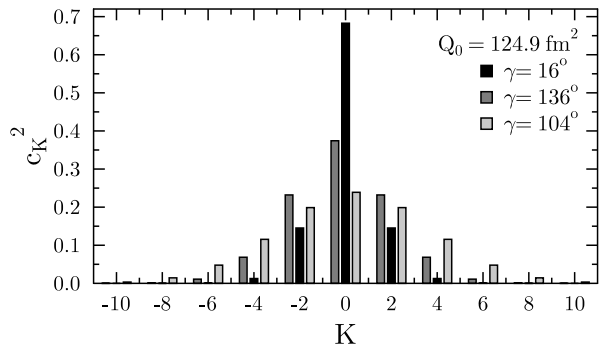}
\caption{\label{fig:decomp:k}
Decomposition of the mean-field state that gives the minimum of the
$J=0$ projected energy surface into components of same $K$,
Eq.~(\ref{eq:cK2}), for three different orientations of the state
in the intrinsic frame.
}
\end{figure}

Figure~\ref{fig:decomp:j} presents the decomposition of the norm
of the same states as in Fig.~\ref{fig:orientation} in components
with different total angular momenta $J$,
\begin{equation}
\label{eq:cJ2}
c_J^2
= \sum_{K = -J}^{+J} \langle q | \hat{P}^J_{KK} | q \rangle
\end{equation}
summed over all possible values of $K$; Fig.~\ref{fig:decomp:k}
shows the decomposition into components with different intrinsic
angular momentum projection $K$,
\begin{equation}
\label{eq:cK2}
c_K^2
= \sum_{J \geq |K|} \langle q | \hat{P}^J_{KK} | q \rangle
\end{equation}
summed over all possible values of $J$ for a given value of $K$.
The underlying state $| q \rangle$ is assumed to be
particle-number projected and normalized. As we always choose an
orientation of the triaxial state $| q \rangle$ where the
intrinsic $z$ axis coincides with one of the principal axes, the
components of opposite $K$ have the same weight $c_{+K}^2 =
c_{-K}^2$. Furthermore, all components with an odd value
of $K$ are zero for all values of $J$.

The decomposition of a triaxial mean-field state into components
with different $J$ values in the laboratory frame should be
independent of its orientation in the intrinsic frame, which is
indeed the case within the resolution of Fig.~\ref{fig:decomp:j}.
The plot suggests a separation of the coefficient $c_J^2$ into two
distinct curves: one for even values of $J$, which dominates the
decomposition and peaks for $J=2$, and a second much weaker one
for odd values of $J$, which peaks at $J=3$. (When decomposing a
time-reversal invariant state, there is no component with $J=1$
for symmetry reasons.) Increasing the deformation of the intrinsic
state modifies the $c_J^2$ distribution such that the peaks of the
distributions for even and odd $J$ are shifted toward larger
values of $J$.

By contrast, the decomposition of the same triaxial intrinsic
state into its $K$ components depends strongly on its orientation,
which underlines that $K$ is not an observable quantity. The
distribution of $K$ components is quite narrow when the long axis
of the intrinsic state coincides with the $z$ axis, and becomes
broader when the intermediate or even short axis is chosen instead.
The different distributions in $K$ space indicate that the
numerical convergence of angular-momentum projection is not the
same for all possible orientations of the mean-field state. The
accurate determination of components with large $K$ values
requires many integration points for the Euler angles $\alpha$ and
$\gamma$; hence, it is more favorable to orient the mean-field
wave function with its $z$ axis along the long axis. The sum over
all $c_J^2$ coefficients should be equal to the sum over all the
$c_K^2$ ones and equal to 1. There is, in practice, a slight
numerical deviation of the order of $10^{-5}$, which can be
attributed to high-$J$ and high-$K$ components requiring a higher
number of integration points than have been used here.

%
%
\subsubsection{Decomposition of an oblate HFB state}
\label{sect:decomp:obl}

\begin{figure}[t!]
\includegraphics{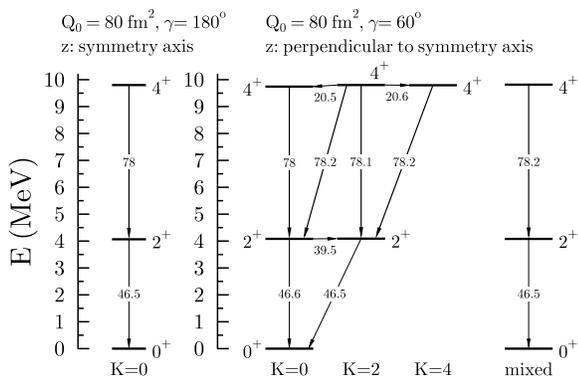}
\caption{\label{fig:oblate}
Excitation spectra and selected $B(E2)$ values after decomposition
of the mean-field state that gives the axial oblate saddle of the
$J=0$ projected energy surface into angular-momentum projected
components $| J M K q \rangle$. The left panel provides the
decomposition when choosing the $z$ axis as the symmetry axis, $\gamma
= 180^\circ$. The middle panel gives the decomposition into $K$
components when the symmetry axis is chosen perpendicular to the
$z$ axis. The right panel shows the unique band resulting from $K$
mixing (see text). }
\end{figure}

A further test of our method is given by the projection of an
oblate mean-field configuration, for which the symmetries of our
codes allow two distinct orientations in its intrinsic frame: The
$z$ axis can be either the symmetry axis or perpendicular
to it. Using the properties of the transformation of the operator
$\hat{P}^J_{MK}$ under rotation, one can show that a pure $K=0$
state is transformed into a multiplet of states with $K$ 
between $0$ and $J$. However, the transformed wave functions
differ only by an unobservable phase and are degenerate. An
example is given in Fig.~\ref{fig:oblate} for the mean-field state
with mass quadrupole moment $Q_0 = 80$ fm$^2$ which gives the axial
oblate saddle point of the $J=0$ projected energy surface. When
using the $z$ axis as symmetry axis,as was done in axial
calculations
\cite{Val00a,Ben03b,Ben04a,Rod02a,Rod02b,Egi04a,Nik06a,Nik06b},
the projection decomposes the mean-field state into a single
rotational band of $K=0$ states; all other $K$ components have
zero norm. When orienting the intrinsic state such that the $z$
axis is perpendicular to the symmetry axis, as done at $\gamma =
60^\circ$ in the $\beta$-$\gamma$ plane, all even $K$ components
$| J M K q \rangle$ up to $K=J$ are nonzero for each $J$. When
constructing the $| J M \kappa q \rangle$ states, however, the
diagonalization of the norm matrix gives only one nonzero
eigenvalue per even $J$, and one ends up with the same rotational
band as obtained by exploiting the symmetry of the intrinsic state.

%
%

\subsubsection{Comparison with the asymmetric rotor model}

\begin{table}[t!]
\caption{\label{tab:asyrot}
Spectroscopic quadrupole moments obtained from projection without
and with $K$ mixing compared with values obtained from the asymmetric
rotor model. In this latter case, the intrinsic charge quadrupole moment
$Q_{0,p} = 63.45$ $e$ fm$^2$ ($\beta_p = 0.583$) and $\gamma = 16.1^\circ$
value of the triaxial HFB state are used as input.
}
\begin{tabular}{lrrr}
\hline\noalign{\smallskip}
Quantity                                       & No mixing & $K$ mixing & Rotor \\
\noalign{\smallskip} \hline \noalign{\smallskip}
$Q_s$ ($2^+_1$)  ($e$ fm$^2$)                  & $-19.1$ & $-18.7$ & $-17.0$ \\
$Q_s$ ($2^+_2$)  ($e$ fm$^2$)                  &   20.0  &   18.1  &   17.0  \\
$B(E2,2^+_1 \to 0^+_1)$ ($e^2$ fm$^4$)         &   94.9  &   95.7  &   75.5  \\
$B(E2,2^+_2 \to 0^+_1)$ ($e^2$ fm$^4$)         &   15.9  &    5.2  &    4.6  \\
$B(E2,2^+_2 \to 2^+_1)$ ($e^2$ fm$^4$)         &    8.4  &   26.4  &   14.3 \\
$B(E2,3^+_1 \to 2^+_1)$ ($e^2$ fm$^4$)         &   11.6  &   10.0  &    8.3 \\
$B(E2,3^+_1 \to 2^+_2)$ ($e^2$ fm$^4$) \qquad  &  146.4  &  165.4  &  134.7 \\
\noalign{\smallskip}\hline
\end{tabular}
\end{table}

The spectroscopic quadrupole moments and $B(E2)$ values obtained
by the projection of a triaxial mean-field state are mainly
determined by its geometry. This property is illustrated in
Table~\ref{tab:asyrot} where we compare the values obtained by
projecting the same triaxial mean-field state as in
Fig.~\ref{fig:orientation} to those calculated with the asymmetric
rotor model introduced by Davydov and co-workers~\cite{Dav58a,Dav59a},
using the intrinsic proton quadrupole moment $Q_{0,p} = 63.45$ $e$
fm$^2$ ($\beta_p = 0.583$) and $\gamma = 16.1^\circ$ as input. The
agreement is excellent and, in practice, improves with
deformation.

%
%

\subsection{Configuration-mixing calculations}

\subsubsection{Selection of the mean-field basis}

As a last step, we perform a mixing of states obtained by
projecting on particle-number and angular-momentum mean-field wave
functions covering the full $\beta$-$\gamma$ plane. Specifically,
the results will be analyzed by comparing the spectra and
transition probabilities obtained in calculations using four
different subspaces of states
\begin{itemize}
\item
a basis of "prolate" axial states, comprising the four
deformations $(q_1, q_2) = (80,0)$, (120,0), (160,0) and (200,0)
fm$^2$;
\item
a basis of "axial" states, where two oblate deformations $(q_1,
q_2) = (0,80)$ and (0,120) fm$^2$ are added to the prolate basis;
\item
a basis of six "triaxial" mean-field configurations $(q_1, q_2) = (80,40)$,
(120,40), (160,40), (200,40), (80,80) and (120,80) fm$^2$;
\item
a basis labeled "full" where, depending on $J$, two to four
states of the prolate basis are added to the triaxial basis.
\end{itemize}
For the full basis, we have added for each $J$ value the largest
possible number of axial points to the triaxial basis.
We have excluded from the GCM calculations those deformed
states that are situated in a region affected by the difficulties
mentioned in Sec.~\ref{sect:MREDF}. This region spreads from
small deformations around the spherical point to a region with
larger deformations between the oblate axis and $\gamma \approx
50^\circ$. This restriction does not permit the mixing of triaxial and
oblate axial states.

\begin{figure}[b!]
\includegraphics{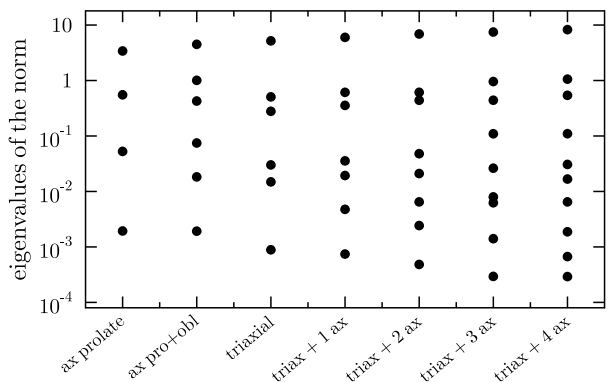}
\caption{\label{fig:norm}
Variation of the eigenvalues of the $J=0$ norm kernel among
different bases.
}
\end{figure}

For \nuc{24}{Mg}, the number of axial states that can be added to
the set of triaxial states is not large, just two to four,
depending on the value for $J$. Redundancies appear  very quickly
in the norm kernel when more states are added to the
nonorthogonal basis. This very small number of nonredundant
states is a direct consequence of the very few level crossings
visible in Fig.~\ref{fig:nilsson:ax}; even states with quite
different intrinsic deformation might have a very similar
single-particle spectrum. This feature is illustrated in
Fig.~\ref{fig:norm}, where the eigenvalues of the $J=0$ norm
kernel are plotted as a function of the number of states included
in the configuration-mixing calculation. We start on the left of
the figure with the eigenvalues of calculations performed in the
prolate and axial bases. The addition of oblate configurations to
the prolate set brings one large eigenvalue, close to 1, and
another one around $10^{-2}$. The range of values obtained for a
purely triaxial set is very similar to the axial set, although
both bases have no vectors in common. Starting from the initial
set of triaxial points, prolate points are added one after the
other. There is, thus, one more eigenvalue after each addition and
the trace of the norm, which is equal to the number of
discretization points, is increased by 1. One can see that this
increase of the trace is mainly distributed among the largest
eigenvalues, which increase slightly. Each time, a new eigenvalue
around $5 \times 10^{-3}$ also appears, or even an smaller one when 
the fourth axial point is added. Although these changes are very small, the
effect of the coupling between axial and triaxial states on all
observables is not completely negligible.

%
%

\subsubsection{Ground-state correlation energy}

\begin{figure}[t!]
\includegraphics{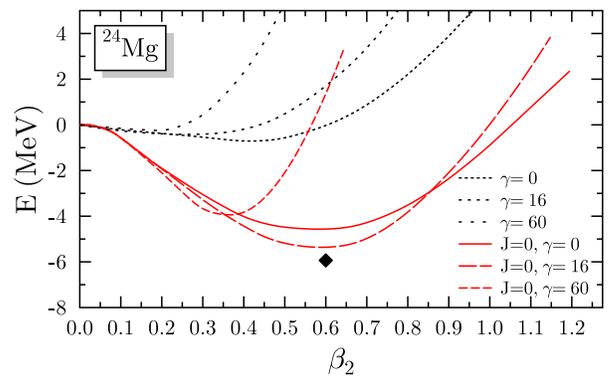}
\caption{\label{fig:ecorr}
(Color online) Deformation energy curve projected on $N$ and $Z$
(black) only and projected also on $J=0$ (red) for three different
values of $\gamma$, as indicated. The energies of the $J=0$ GCM
ground state obtained from the axial, triaxial, and full bases as
described in the text cannot be distinguished within the
resolution of the plot and are represented by the same filled
diamond plotted at arbitrary deformation. All energies are
normalized to that of the spherical particle-number projected
state. }
\end{figure}

We have shown in Fig.~\ref{fig:map:J00} that the angular-momentum
projection changes the topography of the $J=0$ deformation energy
surface and generates a minimum for the projection of a triaxial
configuration. This result is illustrated further in
Figure~\ref{fig:ecorr} where the variation of the energy along
three cuts in the $\beta$-$\gamma$ plane is plotted. The first
curve corresponds to prolate deformations, $\gamma = 0^\circ$, the
second to oblate ones, $\gamma = 60^\circ$, and the third to a cut
along $\gamma = 16^\circ$. Results obtained with and without
projection on $J=0$ are given. The big diamond covers the range of
the GCM energies obtained by using the axial, triaxial, and full
bases. The lowest energy corresponds to a prolate configuration in
the nonprojected calculation, with triaxial energies always much
larger than the prolate and oblate ones. After projection on
$J=0$, the triaxial curve is below the prolate one for a large
range of deformations and the absolute minimum corresponds to a
triaxial configuration about 800 keV more bound than the axial
saddle point. However, the total energy gain obtained by mixing
axial configurations is larger than that from the projection of
a single triaxial configuration. Moreover, there is only a 35 keV
difference between the energies obtained by the mixing of triaxial
configurations and a further gain of 160~keV in the largest
possible set of axial and triaxial configurations.

\begin{figure*}[t!]
\includegraphics{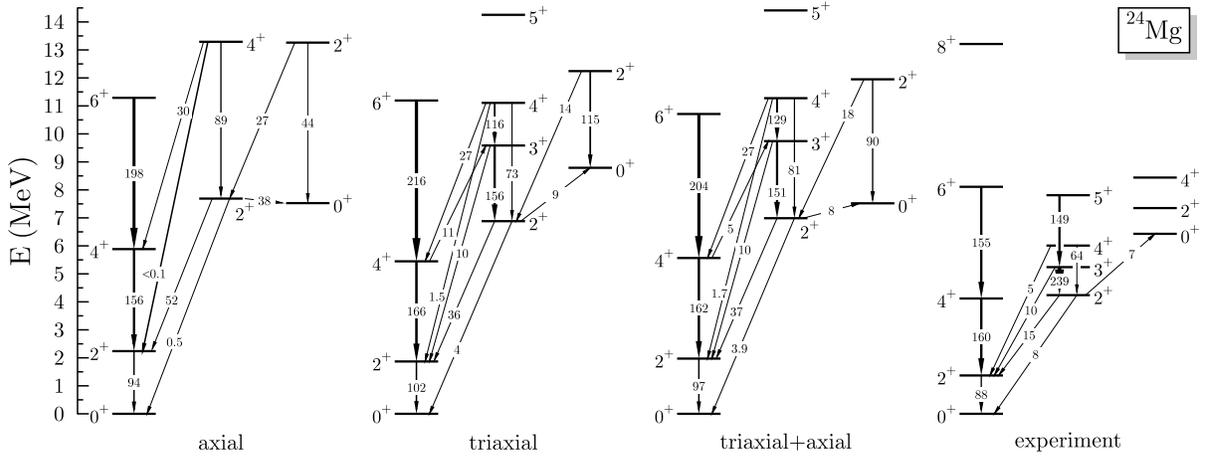}
\caption{\label{fig:spectra}
Excitation spectra and $B(E2)$ values in $e^2$ fm$^4$ obtained in three of the
bases defined in the text compared to the available experimental data.
In the four cases, the spectrum is subdivided into a ground-state band, 
a $\gamma$ band, and additional low-lying states
that do not necessarily form a band.
}
\end{figure*}

This result puts a limit on the meaning of "rotational" and
"vibrational" energies in the ground state: What appears to be
"vibrational" energy in a projected axial quadrupole GCM is
transformed into "rotational" energy in a projected GCM including
triaxial states. At the same time, this result is rather
encouraging as it brings the hope that the ground-state
correlation energies are rapidly saturating when more than the 
restoration of symmetries on the axial mode is included. Hagino \etal\
\cite{Hag03a} arrived at a similar conclusion while studying the
ground-state correlation energy in an exactly solvable model that
simulates collective quadrupole motion. Of course, systematic
realistic calculations, in particular for heavier nuclei, are
needed before a general rule can be established.

%
%
\subsubsection{Excitation spectrum}

The excitation spectra obtained in three different GCM bases are
given in Fig.~\ref{fig:spectra} and Table~\ref{tab:specmom}.
Table~\ref{tab:specmom} also lists results for spectroscopic
quadrupole moments, while the $B(E2)$ values for the most important
transitions are given in Fig.~\ref{fig:spectra}.
Experimental data are taken from Ref.~\cite{Fir07a} in both cases,
using $5.940 \times 10^{-2} \, A^{4/3} \; e$ fm$^4 = 4.11 \; e$ fm$^4$
for the $E2$ Weisskopf unit.

The spacing of levels, the relative strength of $B(E2)$ values and
the $K$ content of the collective wave functions suggest a separation
of the low-lying spectrum into three bands: The $\Delta J = 2$ ground-state
band dominated by $K=0$ components, a $\Delta J = 1$ $\gamma$ band
dominated by $K=2$ components, and a third band that is again dominated
by $K=0$ components.

\begin{table}[b]
\caption{\label{tab:specmom}
Comparison between the theoretical excitation energies in MeV and
spectroscopic quadrupole moments in units of $e$ fm$^2$
and the experimental  values taken from Ref.~\cite{Fir07a}.
}
\begin{tabular}{lccccccccc}
\hline\noalign{\smallskip}
Level &
\multicolumn{4}{c}{$E_{ex}$} & &
\multicolumn{4}{c}{$Q_s$}  \\
\noalign{\smallskip} \cline{2-5} \cline{7-10}\noalign{\smallskip}
         & Axial  & Triaxial & Full & Expt. &  & Axial & Triaxil & Full & Expt. \\
\noalign{\smallskip} \hline \noalign{\smallskip}

$2^+_1$  &  2.24 &  1.87 &  1.97 & 1.37 & & $-17.1$ & $-19.6$ & $-19.4$ & $-16.6$ (6) \\
$4^+_1$  &  5.88 &  5.44 &  5.57 & 4.12 & & $-25.1$ & $-26.1$ & $-26.0$ &             \\
$2^+_2$  &  7.69 &  6.88 &  6.99 & 4.24 & &    9.9  &   17.1  &   16.6  &             \\
$3^+_1$  &  ---  &  9.59 &  9.74 & 5.24 & &   ---   &  $-0.1$ &  $-0.1$ &             \\
$4^+_2$  & 13.29 & 11.12 & 11.28 & 6.01 & &    9.0  &  $-7.3$ &  $-7.4$ &             \\
$0^+_2$  &  7.53 &  8.79 & 7.520 & 6.42 & &    0.0  &    0.0  &    0.0  &    0.0      \\
\noalign{\smallskip}\hline
\end{tabular}
\end{table}

As already mentioned, the GCM ground-state energy is quite
close in the purely axial and purely triaxial calculations, but is
160~keV lower in the full calculation that combines triaxial and
axial states. All excited levels except the $0^+_2$ one, are lower
in the triaxial basis than in the axial one. Adding prolate axial
states to the triaxial basis mainly lowers the energy of all $0^+$
levels, pushing up levels with other values of $J$. The
spectroscopic quadrupole moments $Q_s$ and $B(E2)$ values,
however, are not significantly affected.

The most significant difference between the axial and triaxial
calculations concerns the first excited band, which is clearly a
$\gamma$ band in the triaxial basis. Odd-$J$ levels are of course
absent from the first excited band in the axial basis but some of
the features of a $\gamma$ band are already hinted in this band
dominated by oblate configurations.  In particular, the $B(E2)$
value for the transition from the $4^+_2$ to the $2^+_2$ level is
very close to that of the triaxial calculation. This
identification of the projection of an oblate state to a $\gamma$
band was already suggested in  Sec.~\ref{sect:decomp:obl}.
However, $K$ mixing does not bring any gain in energy in the
projection of an oblate state since the relative weights of the
$K$ components are fixed by symmetry relations, whereas they are
free to vary in the projection of triaxial states. The gain in
energy brought by the inclusion of triaxial configurations is
evident when looking at the energies and the spectroscopic
quadrupole moment (Table~\ref{tab:specmom}), which change
drastically when going from the axial to the triaxial calculation.

The inclusion of triaxial deformations has also a strong effect on
the energies of the levels in the second band. The excitation
energy of the $2^+_2$ level is lowered by nearly 1~MeV, and that
of the excitation energy of the $4^+_2$ level even more.
Altogether, the inclusion of triaxial mean-field states brings the
right tendency to make the spectrum more compact and brings it
closer to the experimental data. It remains to verify whether the
use of time-reversal invariance breaking HFB states will add the
extra gain of energy required to bring the states with $J$
different from $0$ closer to the experimental levels. In
particular, the odd-$J$ states in the $\gamma$ band still have
excitation energies that are much too high.

\begin{table}[t!]
\caption{\label{tab:overlap_gcm}
Overlaps between the GCM wave functions obtained in calculations
using different sets of mean-field states. The two results marked
by asterisks indicate that the number given represents the overlap
between the second prolate and the third triaxial or full $2^+$
excited states.
}
\begin{tabular}{lccccc}
\hline\noalign{\smallskip}
        & \multicolumn{2}{c}{Triaxial} & & \multicolumn{2}{c}{Full} \\
\noalign{\smallskip}\cline{2-3} \cline{5-6}\noalign{\smallskip}
        & Prolate & Axial & & Prolate & Axial \\
\noalign{\smallskip} \hline \noalign{\smallskip}
        & 0.97     & 0.99 & & 0.97     & 0.99 \\
$J=0$   & 0.89     & 0.94 & & 0.96     & 0.97 \\
        & 0.16     & 0.93 & &  --      & 0.96 \\
\noalign{\smallskip} \hline \noalign{\smallskip}
        & 0.97     & 0.98 & & 0.97     & 0.98 \\
$J=2$   & 0.21     & 0.86 & & 0.23     & 0.88 \\
        & 0.88$^*$ & 0.82 & & 0.88$^*$ & 0.81 \\
\noalign{\smallskip} \hline
\end{tabular}
\end{table}

\begin{table}[b]
\caption{\label{tab:transmom}
Comparison between theoretical and experimental $B(M1)$ values in
units of $\mu_N^2$. Data taken from Ref.~\cite{Fir07a}.
}
\begin{tabular}{lcccc}
\hline\noalign{\smallskip}
transition &  \multicolumn{4}{c}{$B(M1)$} \\
\noalign{\smallskip} \cline{2-5}  \noalign{\smallskip}
           &  Axial & Triaxial & Full & Expt.  \\
\noalign{\smallskip} \hline \noalign{\smallskip}
$2^+_2 \to 2^+_1$  &$ 4 \times 10^{-6}$ & $3 \times 10^{-6}$ & $3 \times 10^{-6}$ & $1.6 \times 10^{-5}$ (14) \\
$3^+_1 \to 2^+_1$  &    ---            & $2 \times 10^{-7}$ & $1 \times 10^{-7}$ & $3.8 \times 10^{-5}$ (20) \\
$3^+_1 \to 2^+_2$  &    ---            & $4 \times 10^{-5}$ & $4 \times 10^{-5}$ & $6.2 \times 10^{-4}$ (30) \\
\noalign{\smallskip}\hline
\end{tabular}
\end{table}

To analyze further the equivalence of and differences between the
bases that we have used, the component of the triaxial and full
bases that are included in the prolate and axial bases are given
in Table~\ref{tab:overlap_gcm}. These overlaps, $\langle J M \nu |
J M \mu \rangle$, can be easily calculated by using the $F^J_\nu
(\kappa,q)$ and the norm kernels
$\mathcal{I}(\kappa,q;\kappa',q')$. A bit surprisingly, the
differences between the collective ground-state wave functions
obtained within the axial and full bases are very small, the
overlap with the prolate basis being slightly lower but still
quite high. The second $0^+$ state is not as well described by the
prolate basis and the third one state is missing in this basis.
Note that the  difference between the axial and full ground-state
wave functions is still larger than the energy differences between
these states (0.16~MeV out of a total energy of around 200~MeV).

The first $2^+$ state has the same structure in all bases. In
contrast, the second $2^+$ state of the axial and full bases is
not described by the prolate basis. Although the overlap
between the $2^+_2$ states in the axial and full bases is close to
0.9, the excitation energy is lowered by 800~keV when triaxial
configurations are included. This confirms our previous interpretation 
that the mixing of prolate and oblate axial states can, to
some extent, simulate states with $K \neq 0$, but not fully.

In the \mbox{$N=Z$} nucleus \nuc{24}{Mg}, all calculated magnetic
moments are just a few percent larger than $Z J/(N+Z) = J/2$ in
nuclear magnetons. This is a consequence of (a) the time-reversal
symmetry that we impose on the underlying HFB states as explained
in Sec.~\ref{sect:trans} and of (b) the fact that protons and
neutrons have nearly the same contribution to the angular momentum
in this $N=Z$ nucleus. The calculated magnetic moments agree well
with the available experimental ones~\cite{Fir07a} for the
$2^+_1$, $4^+_1$, $2^+_2$, and $4^+_2$ states within the experimental
error bars.

The calculated and experimental $B(M1)$ values are compared in
Table \ref{tab:transmom}. Data are taken from Ref.~\cite{Fir07a}
where $1.790 \, \mu_N^2$ is used for the $M1$ Weisskopf unit.
The values that we obtain are about one order of magnitude too
small, which clearly indicates that the projected currents and spin
densities are underestimated. Starting from time-reversal breaking
mean-field states instead of time-reversal invariant ones as done
here can be expected to increase the $B(M1)$ values.
%
%
\section{Summary and Outlook}

A generalization of a method that enables the mixing of projected
mean-field states that was previously limited to axial configurations
has been set up to allow for a description of the full
five-dimension quadrupole dynamics. Compared to a GCM calculation
limited to projected axial quadrupole deformed states, the present
method allows for a spreading of the collective  states into the
$\beta$-$\gamma$ plane. In the case of \nuc{24}{Mg}, we have shown
how the spectroscopic properties of the low-lying states are
affected. One can summarize our main results as follows:
\begin{itemize}
\item
When looking at the projection of a single configuration, the
energy obtained for the ground state is significantly lowered when
allowing for triaxial quadrupole deformation;
\item
If one considers at the same time the correlation energy from
symmetry restorations and configuration mixing, the total
energy difference between the ground state obtained within an
axial and a triaxial mean-field basis is quite small. This
indicates that the nondiagonal matrix elements between prolate
and oblate axially deformed mean-field states bring a large
fraction of the correlation energy that is obtained by the
projection of triaxial configurations between them. This puts a
limit on the meaning of rotational and vibrational energies, as
what appears to be "vibrational" energy in the projected
quadrupole GCM  of axial states is transformed into "rotational"
energy in a projected GCM including triaxial states. Of course,
this result will have to be confirmed by studies in other nuclei;
\item
This finding is also supported by the analysis of the spectrum of
eigenvalues of the norm matrix, where adding triaxial states does
not introduce states with large eigenvalues to the space of $0^+$
states.
\item
For higher $J$ values, the situation is more complex. However,
including triaxial deformations lowers the excitation energies and
brings the spectrum closer to experiment.
\end{itemize}
On the basis of this analysis, one can draw some conclusions about
the effect of triaxial deformations:
\begin{itemize}
\item
Their effect on masses seems marginal and it is reassuring in some
way: If one can confirm that triaxial deformations increase binding
energies by only around 100 to 200~keV, it would be justified to avoid the
complexities from their introduction in systematic mass calculations.
\item
The gain of energy on excited states partly cures a problem common to all
projected GCM calculations based exclusively on axial mean-field
states. However, further improvements are still necessary, including
the breaking of time-reversal invariance and the
consideration of the projection of cranking states optimized for each
angular momentum. The breaking of axiality is a  necessary first step
before the breaking of time-reversal invariance and we hope to validate
the extension of our method in this case in the near future.
\item
The power of our method will be more apparent when breaking of time-reversal
invariance is included. Although some new states can already be described
at the present level of development (e.g., odd-$J$ members of $K\neq 0$ bands),
it will be possible to describe quasiparticle excitations and in particular
nuclei with an odd number of neutrons or protons.
\item
Our method also provides an ideal tool to benchmark simpler models. To give
only one example, the metric of the $\beta$-$\gamma$-plane is generated
directly in our method by the overlaps between mean-field wave functions
of different shapes. Although it is not trivial to derive this metric in
a multidimensional problem, as was done in Ref.~\cite{Bon90a} for the
one-dimensional case, it would be  very instructive to compare a metric
derived from a purely microscopic approach to the metrics that are
usually based on semiclassical approximations.
\end{itemize}
There are no basic reasons that prevent the application of the
method presented here to heavy nuclei, although this would be too 
time consuming with our current numerical implementation. Further
developments are required to improve the efficiency of the
numerical algorithms. In particular, the choice of discretization
points used to evaluate the integrals over the Euler angles
needs to be optimized and the codes have to be parallelized, which
could be done in a very efficient way. Work in that direction is
underway.

Another development that will have to be completed in the near
future is the implementation of the regularization scheme proposed
in Ref.~\cite{Lac08a} to remove the pathologies brought by the use
of the generalized Wick theorem when evaluating the energy kernels
starting from an  energy density functional. However, the
conceptual and technical difficulties encountered in the present
generalization of the projected GCM justify the continuation of the 
present developments in parallel. Considerable work at the level
of the computational algorithm still remains to be performed to
have a method applicable to heavy nuclei.

%
%
\section*{Acknowledgments}
We thank P.~Bonche and J.~Dobaczewski for many valuable and
enlightening discussions on various aspects of this work and
R.~V.~F.~Janssens for a critical reading of this manuscript. This research
was supported in parts by the PAI-P5-07 of the Belgian Office for
Scientific Policy, by the U.S.\ Department of Energy under Grant Nos.\
DE-FG02-00ER41132 (Institute for Nuclear Theory) and 
DE-AC0Z-06CH11357 (Argonne National Laboratory), and by the U.S.
National Science Foundation under Grant No.\ PHY-0456903 (MSU).
Part of the work of M.~B.\ was performed within the framework of
the Espace de Structure Nucl{\'e}aire Th{\'e}orique (ESNT). M.~B.\
also acknowledges support from the European Commission during the
initial stage of this work.
%
%
\begin{appendix}

\section*{Appendix: Comments on the numerical evaluation of the kernels}
\label{app:representation}

\subsection{Self-consistent mean-field calculations}

The mean-field states are generated with a simplified version of
the cranked HFB code whose evolution is documented in
Refs.~\cite{Bon87a,Gal94a,Rig99a}. We have imposed time-reversal
invariance of the HFB vacua such that the single-particle wave
functions are pairwise connected by time reversal. The
single-particle wave functions are represented as complex spinors
discretized on a three-dimensional Cartesian Lagrange mesh in
coordinate space. In addition to time reversal, two further
symmetries of the $D_{2h}^{TD}$ symmetry
group~\cite{Dob00a,Dob00b} are imposed on the single-particle
basis, namely that they are eigenstates of the parity $\hat{P}$
and $z$ signature $\hat{R}_z$. Their relative phases are fixed by
choosing a basis where the eigenvalue of the $y$ time simplex
$S^T_y$ is $+1$ for all single-particle states. This introduces
three plane symmetries and allows us to restrict the numerical
representation of individual wave functions to 1/8 of a full
box~\cite{Bon87a}. In addition, the single-particle states are
chosen to have good isospin projection (i.e., they are either pure
proton or neutron states).

Mean-field states with different deformation are obtained adding
constraints on $q_1$ and $q_2$ as defined in
Eq.~(\ref{eq:constraint}) in the variation. The constrained HFB
equations are solved by using the "two-basis method" described in
Ref.~\cite{Gal94a}, which delivers the HFB states $| q \rangle$
through quite a small number of single-particle states represented
in coordinate space in the so-called Hartree-Fock (HF) basis that
diagonalizes the single-particle Hamiltonian and the corresponding
$U$ and $V$ matrices that establish the general Bogoliubov
transformation \cite{Rin80a,BlaRip,Man75a} from the HF basis to
the quasiparticle basis that diagonalizes the quasiparticle
Hamiltonian. This procedure permits to limit the numerical
representation to all single-particle levels below the Fermi
energy and to a small number of levels above. As done in our
earlier configuration-mixing calculations, we add the Lipkin-Nogami 
(LN) prescription to the HFB equations to enforce pairing correlations
in all mean-field states. Using states without pairing in a GCM
calculation introduces the danger of artificially decoupling
many-body states with a different ratio of occupied
single-particle states of positive and negative parity, which can
lead to instabilities when solving the HWG
equation~(\ref{eq:HWG:full}).

The representation of the single-particle states on a coordinate space
mesh has the clear advantage that its precision is fairly independent
of the deformation when a sufficiently large box is chosen, and it only
depends on the distance of discretization points. This is important
for GCM calculations mixing states with very different deformation.

For the subsequent projection and mixing of HFB states with
different shapes, it is advantageous to use the canonical
single-particle bases of the mean-field states as starting point.
This simplifies the corresponding $U$ and $V$ matrices and allows
for a safe cutoff of single-particle states with very small
occupation $v^2$ that do not contribute to any of the kernels. For
the identification of symmetries of the integrals over Euler
angles, it also turns out to be advantageous to transform the
single-particle basis to a basis of eigenstates of $x$ signature
$\hat{R}_x$.
%
%

\subsection{Rotation of mean-field states}

The rotation of the "left" state in coordinate space and the "right"
state in gauge space, Eq.~(\ref{eq:meLR}), can be performed either
as rotations of the
canonical single-particle states leaving the corresponding $U$ and
$V$ matrices untouched or as rotations of the $U$ and $V$
matrices leaving the single-particle states untouched. For the
coordinate space rotation, the latter is equivalent to the
expansion of the rotated single-particle states in terms of the
unrotated ones, which already at moderate deformation requires
highly excited single-particle states above the Fermi energy, which
are outside of the single-particle basis used to describe the
unrotated state. Therefore, the coordinate space rotation
$\hat{R}(\alpha,\beta,\gamma)$ is performed as a rotation of the
single-particle states on the mesh as described in
Refs.~\cite{Bay84a,Bay86a}, which, however, is the most
time-consuming piece of the numerical calculations. In contrast,
it is simpler and numerically more efficient to perform the gauge-space 
rotation as a transformation of the $V$ matrices instead of
the single-particle basis, which in the canonical basis boils down
to the multiplication of a small antidiagonal matrix with a phase
factor.

The rotation operator $\hat{R}(\alpha,\beta,\gamma)$ mixes
single-particle states of both signatures, which requires one to
extend the numerical representation of the single-particle wave
functions from 1/8 to 1/2 of the full box, leaving only parity
(and the isospin projection) as a good quantum number. The
symmetries of the unrotated time-reversal invariant HFB states,
however, permit the reduction of the integration intervals for
Euler angles to $\alpha \in [0,\pi/2]$, $\beta \in [0,\pi/2]$, and 
$\gamma \in [0,\pi]$ (i.e., 1/16 of the full $8\pi^2$ integration
volume for systems with integer spin).

%
%

\subsection{Calculation of the kernels between rotated mean-field states}

Rotating an HFB state in coordinate or gauge space gives back an HFB
state; hence, the matrix elements between "left" and "right"
states, Eq.~(\ref{eq:meLR}), can be easily evaluated with the
generalized Wick theorem \cite{Bal69a,Rin80a}. The expressions
given in Refs.~\cite{Bal69a,Rin80a} cannot be directly used,
however. First, we have to transform the contractions between
states in different quasiparticle bases to expressions for
contraction between states in different canonical single-particle
bases \cite{Bon90a,Hee93a}. Second, our coordinate space
representation has as a consequence that the single-particle bases
that set up $| q \rangle$ and $| q' \rangle$ are not equivalent, a
difficulty that is amplified further by rotation of one of the
states. This difficulty can be overcome by eliminating the
contribution to the kernels coming from single-particle states
that are occupied  in one of the bases but not in the other
\footnote{
The expressions for the matrices defining the basic contractions
and the overlap given in Refs.~\cite{Bon90a,Hee93a,Val00a}
contain a  systematical typographical error: all
$(\mathcal{R}^\dagger)^{-1}$ should be replaced by
$(\mathcal{R}^T)^{-1}$, which does not have any consequences
for these papers as the matrix $\mathcal{R}$ is real
with the symmetries chosen there.
}
\cite{FloThesis,Bon90a,Hee93a,Val00a}.

Only diagonal kernels and half of the off-diagonal ones have to be
calculated; the other half of the off-diagonal kernels can be
constructed by using symmetries of the kernels.
%
%

\subsection{Particle-number projection and the phase of the overlap}

The integrals over the gauge angles for projection on proton and
neutron number are discretized with the Fomenko prescription
\cite{Fom70a}, which is equivalent to a trapezoidal rule. By using the
number parity of the mean-field states, the integration intervals
can be reduced to $\phi \in [0,\pi]$ for protons and neutrons.
Additionally, a symmetry connects the basic contractions and
overlap in the interval $[0,\pi/2]$ with those in $[\pi/2,\pi]$,
which, however, cannot be used to reduce the integration intervals
for the reasons outlined in the Appendix of Ref.~\cite{Hee93a}.

With the symmetries of the HFB states chosen here, the overlap of
two HFB states with different deformation, where one is
additionally rotated in coordinate space, remains real. As soon as
one of the two HFB states is additionally rotated in gauge space,
however, the overlap in general becomes complex. Its modulus is
determined by the Onishi formula \cite{Oni66a,Rin80a}. Its phase,
a rapidly varying oscillatory function of the gauge angle of
particle-number projection, has to be determined from continuity
arguments. Its value is zero for the overlap between
non-gauge-space-rotated states and is followed during the
gauge-space rotation by performing a second-order Taylor
expansion. To this end, the overlap and its derivatives are also
determined at a small number of gauge angles between the
integration points for the gauge-space integration, separately for
protons and neutrons.

\end{appendix}
%
%

\end{document}